\documentclass[aps,prl,twocolumn,groupedaddress,showpacs]{revtex4}

\usepackage{graphicx}

\begin{document}

\title{Pressure induced variation of the ground state of CeAl$_3$}
\author{M. Weller$^{1}$}
\email[]{weller@phys.ethz.ch}
\homepage[]{www.solid.phys.ethz.ch/ott}
\author{J. L. Gavilano$^{2,1}$, A. Sacchetti$^{1}$, and H. R. Ott$^{1}$}
\affiliation{$^{1}$Laboratorium f\"ur Festk\"orperphysik, ETH Z\"urich, CH-8093 Z\"urich, Switzerland\\
$^{2}$Laboratory for Neutron Scattering, Paul Scherrer Institut, Switzerland}

\date{\today}

\begin{abstract}
Pressure-induced variations of $^{27}$Al NMR spectra of CeAl$_3$ indicate significant changes in the ground-state characteristics of this prototypical heavy-electron compound. Previously reported magnetic and electronic inhomogeneities at ambient pressure and very low temperatures are removed with external pressures exceeding 1.2~kbar. The spectra and results of corresponding measurements of the NMR spin-lattice relaxation rates indicate a pressure-induced emergence of a simple paramagnetic state involving electrons with moderately enhanced masses and no magnetic order above 65~mK.
\end{abstract}
\pacs{71.27.+a, 75.20.Hr, 76.60.-k}
\keywords{Strongly correlated electrons, CeAl3, NMR, Pressure}

\maketitle

At low temperatures, the specific heat of CeAl$_3$ is dominated by a contribution that varies linearly with temperature. Likewise, a large Pauli type magnetic susceptibility, $\chi(T\rightarrow0)= 0.036$~emu~mol$^{-1}$, indicates a much enhanced density of electron states at the Fermi energy $D(E_F)$, a feature that is typical for heavy-electron compounds~\cite{Andres1975}. The ground state of CeAl$_3$ has been investigated with a large number of experimental techniques. An early indication for inhomogeneities of this ground state was provided by $\mu$SR experiments~\cite{Barth1987, Barth1989} and this conjecture was later confirmed by $^{27}$Al NMR results~\cite{Gavilano1995}. Subsequently, based on NMR results, Nakamura et al.~claimed a magnetically ordered ground state for CeAl$_3$~\cite{Nakamura1988}, while Wong and Clark, again employing NMR, found no evidence for magnetic ordering~\cite{Wong1992}. The early results suggesting an inhomogeneous ground state revealed two phases; one of them exhibiting quasi static magnetic correlations, leading to a distribution of local magnetic fields at the Al sites, and a second, paramagnetic phase. The corresponding NMR experiments were made on a powder sample and  the role of strain effects due to the powdering of the sample material and thus introducing inhomogeneities that mimic an inhomogeneous ground state, remained unclear. NMR experiments under pressure as those presented here aimed at resolving this issue. Experimental data at pressures up to 15~kbar probing the specific heat and the electrical resistivity are available in the literature~\cite{Brodale1986,Kagayama1994}. To our knowledge, no experimental investigation probing microscopic aspects of binary CeAl$_3$ under pressure has been reported so far. Our results reveal that the ground state of CeAl$_3$ is very sensitive to pressure induced volume changes. In particular we note that the two mentioned phases collapse into one at surprisingly moderate pressures.

The $^{27}$Al NMR experiments on CeAl$_3$ at low temperatures and under pressure were made in a top-loading dilution refrigerator. The CeAl$_3$ powder sample was the same that was investigated in the previous NMR study~\cite{Gavilano1995}. A possible deterioration of the sample was checked by repeating the measurement of the NMR spectrum at low temperatures and ambient pressure. No detectable changes of the spectra were observed. For the experiments under pressure, a BeCu piston cylinder type pressure cell with 14 (3.8) mm outer (inner) diameter was used. In the pressure cell, the CeAl$_3$ sample was embedded in paraffin and surrounded by a small NMR rf-coil with a volume of $\sim$1 mm$^3$. The pressure was determined by the $^{63}$Cu nuclear quadrupole resonance frequency of Cu$_2$O at $T=1$~K~\cite{Reyes1992}. The powdered Cu$_2$O, also embedded in paraffin, was placed in a second rf-coil. 
The NMR resonant circuit consisted of the inductance in the pressure cell containing the sample and of two capacitors placed outside the pressure cell. The NMR signal was monitored with the standard spin echo method using a home built heterodyne, phase sensitive NMR spectrometer. The $^{27}$Al NMR spectra were recorded at a fixed frequency, varying the magnetic field. The magnetic field was calibrated using the $^1$H NMR resonance from the protons which are present in the pressure medium~\cite{footnote1} and in the paraffin. This approach assures that possible magnetic field changes due to paramagnetic impurities in the pressure cell material can be taken into account. Monitoring the width of the $^{63}$Cu signal gave evidence that the applied pressure was indeed hydrostatic. The spectral weight (spin echo intensity) was obtained by integrating the detected spin echo signal in the time domain. The spin-lattice relaxation rate (SLRR) $T_1^{-1}$ was measured on the central NMR line with a pulse sequence made up of a comb of 10 pulses (8~$\mu$s rf pulses, 35~$\mu$s time lag), a variable delay and a spin echo sequence.
The main experimental difficulty was to cope with the low measuring frequency of about 1~MHz, resulting in very small signals. This and the long spin-lattice relaxation times at low temperatures provoked measuring times of several days for a single NMR spectrum or a magnetisation recovery curve. In order to minimize undesired effects of external magnetic fields on the ground state of CeAl$_3$, also at elevated pressures, and possible heating effects, we restricted our measurements to Larmor frequencies of the order of 1~MHz.

\begin{figure}
\includegraphics{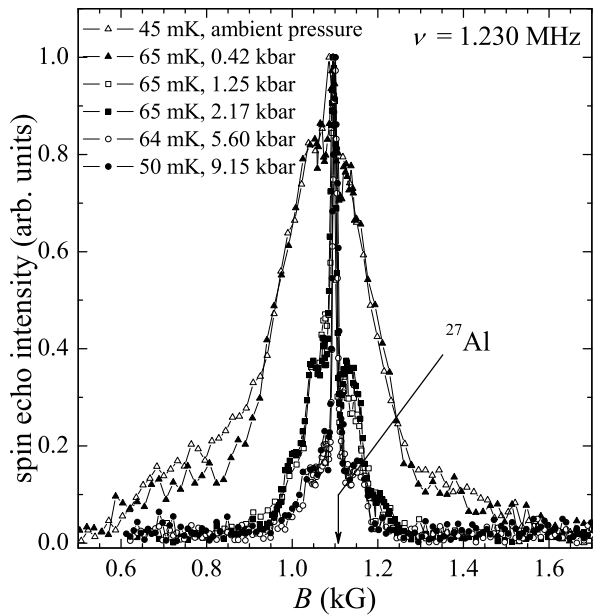}
\caption{$^{27}$Al NMR spectra of CeAl$_3$ for different pressures at very low temperatures. \label{SPC_all}}
\end{figure}
\begin{figure}
\includegraphics[width=7cm]{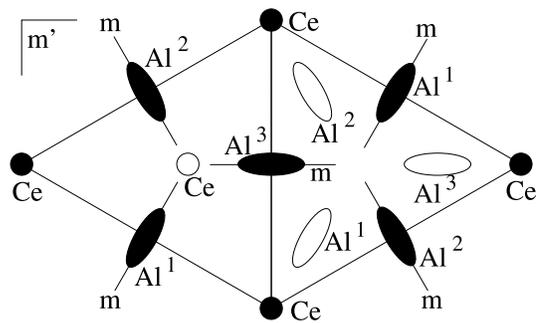}
\caption{Symmetry considerations for $^{27}$Al NMR in CeAl$_3$. The black circles and ellipses denote ion positions, which lie in the same plane perpendicular to the c axis, i.e., the mirror plane $m'$ at $z=1/4$. The open circles and ellipses are sites at $z=3/4$\label{symmetry}}
\end{figure}

In figure~\ref{SPC_all}, $^{27}$Al NMR spectra recorded at $T\approx 65$~mK and at different pressures are displayed.
The ambient pressure results coincide with the data obtained in the quoted former work~\cite{Gavilano1995} on the same sample. A drastic narrowing of the $^{27}$Al NMR spectrum is observed for $p$ exceeding 1.25~kbar. The spectrum then consists of a narrow central line and a broader wing pattern. The spectra recorded at 1.25 and 2.17~kbar are very similar in shape, but still much broader than those obtained at 5.60 or 9.15~kbar.
The central line and the wings are both affected by the Knight-shift and the electric-field gradient. Hence, the precise identification of the relevant parameters directly from the spectra is difficult; the employed procedure is explained below. The crystal structure of CeAl$_3$ has important implications on the $^{27}$Al NMR features. The Al sites are on positions with a point symmetry $mm$ in the hexagonal lattice. The low point symmetry dictates axially nonsymmetric tensors and three Al sites must be distinguished. In figure~\ref{symmetry}, the three different sites are designated as Al$^1$, Al$^2$, and Al$^3$, respectively. The relevant NMR parameters are captured in the tensors for the Knight shift $K$ and the electric field gradient (EFG), respectively. For all three Al sites, the tensors are equal, but rotated by 120 degrees around the c-axis (figure~\ref{symmetry}).
In external magnetic fields with arbitrary orientations, the three Al sites are inequivalent.

\begin{figure}
\includegraphics{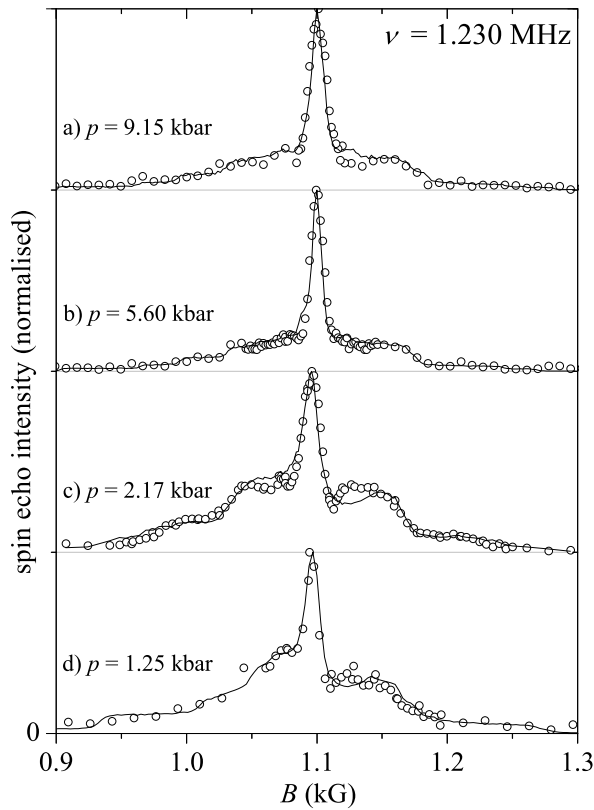}
\caption{Simulated $^{27}$Al NMR spectra (solid lines) of CeAl$_3$ for different pressures. Empty circles represent the same experimental data as in figure~\ref{SPC_all}.\label{Sim}}
\end{figure}

In a powder sample all orientations of the crystalline axes with respect to the magnetic field are equally distributed. The calculation of $^{27}$Al NMR powder spectra can thus be performed assuming only one Al site as long as no magnetic order or any other symmetry breaking is present. These calculations may be based on an effective static nuclear Hamiltonian
\begin{equation}
{\cal H}={\cal H}_{\rm Z}(B_0, K_{\rm iso},K_{\rm ani},\epsilon)+{\cal H}_Q(V_{zz}, \eta)\label{hamiltonian},
\end{equation}
i.e., the sum of a Zeeman term ${\cal H}_{\rm Z}$, including an anisotropic Knight-shift $\tensor{K}$, and a nuclear quadrupolar term ${\cal H}_Q$~\cite{Carter1977}.
The eigenvalues of ${\cal H}$ depend on the orientation of the external magnetic field. The largest components of the Knight-shift tensor $K_Z=K_{\rm iso}+K_{\rm ani}$ and of the electric-field gradient, $V_{zz}$, were previously found to be parallel to the $c$-axis~\cite{Gavilano1995}. Thus the spatial relation between the Knight shift and the EFG tensors is known, except for the relative orientation in the $x-y$ plane, which is restricted by symmetry to two cases with either $V_{xx}\parallel K_X$ or $V_{xx}\parallel K_Y$. It is assumed that these symmetry relations are not altered by the application of moderate hydrostatic pressures.
With these considerations the powder spectra could fairly well be reproduced by simulation calculations, varying the free parameters given in equation \ref{hamiltonian} (see figure~\ref{Sim}). The non-perturbative calculation was based on a full diagonalisation of the static nuclear Hamiltonian and the powder spectrum was simulated with 10'000 arbitrary crystalline orientations. The parameters, which were found to reproduce the spectra are summarised in table~\ref{ceal3_sim_table}~\cite{footnote2}. Our extensive simulations, involving a wide range of parameters, allowed the estimate of statistical uncertainties shown in table~\ref{ceal3_sim_table}. These uncertainties represent the change of the parameters that would increase chi-squared by 50\%. The change in $K_{\rm iso}$ is thus significant, because it varies by a factor 2 over the covered pressure range.
\begin{table}
\centering\begin{tabular}{ccccccc}
\hline\hline
$p$  & $K_{\rm iso}$ & $K_{\rm ani}$  & $\epsilon$ & ~$V_{zz}$~ & $\eta$ & $LW$ \\
(kbar) & (\%) &  (\%) & 1 &  (a.u.)& 1 &  (kHz)\\
\hline
~1.25~ & ~2.1(2)~ & ~3.5(9)~ & ~0~ & ~0.026(3)~ & ~0.40(10)~ & 4 \\
~2.17~ & ~2.1(2)~ & ~3.2(8)~ & ~0~ & ~0.026(3)~ & ~0.15(4) ~ & 6 \\
~5.60~ & ~1.0(1)~ & ~1.0(3)~ & ~0~ & ~0.018(2)~ & ~0.50(13)~ & 2 \\
~9.15~ & ~0.9(1)~ & ~0.8(2)~ & ~0~ & ~0.021(2)~ & ~0.45(11)~ & 3 \\
\hline
ambient & ~2.1(1)~ & 0.8(1)  & ~0~ &  0.019     & 0.2        & - \\
\hline\hline
\end{tabular}
\caption{Summary of the parameters used for the calculated $^{27}$Al NMR spectra. $K_{\rm iso}$, $K_{\rm ani}$ and $\epsilon$ represent the anisotropic Knight-shift, $V_{zz}$ and $\eta$ characterise the electric-field gradient. The parameter $LW$ reflects the broadening of the calculated discrete resonance-fields. The last row cites data from~\cite{Hunziker1996}, evaluated at 35~mK, for comparison. Numbers in parenthesis are error estimates of the respective parameters. \label{ceal3_sim_table}}
\end{table}

\begin{figure}
\includegraphics[width=8cm]{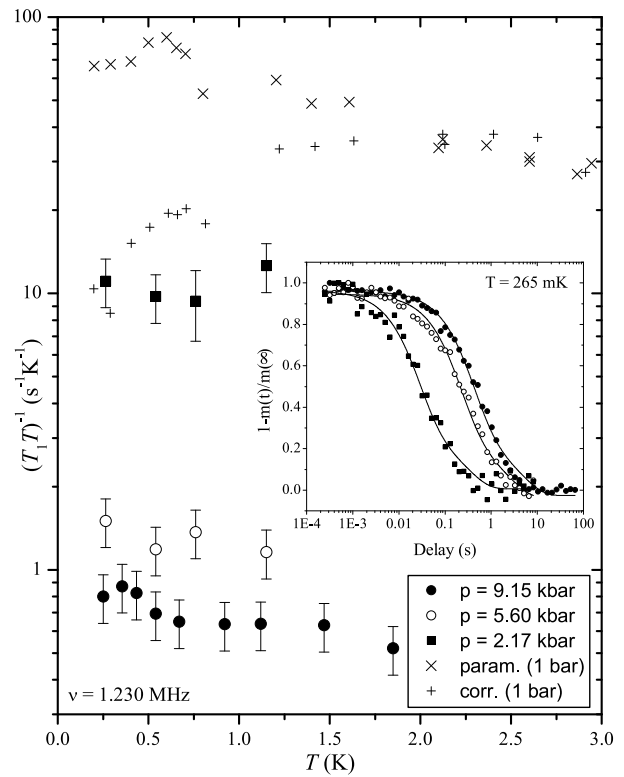}
\caption{$^{27}$Al SLRR of CeAl$_3$ for different pressures and temperatures. We also measured $T_1$ at 1.25~kbar at the lowest temperature. The corresponding value is very close to that measured at 2.17~kbar. Ambient pressure data are taken from~\cite{Gavilano1995}. For LaAl$_3$, the $T$-independent value for $(T_1T)^{-1}=0.037~\rm (sK)^{-1}$~\cite{Joon1989}. \label{T1T}}
\end{figure}

In the experiments probing the SLRR at non zero pressure, we irradiated only the narrow central line in a field range of approximately $\pm30$~G around the maximum~\cite{footnote3}. Therefore, the recovery of the nuclear magnetisation cannot be described by a single exponential and the data need to be fitted to the relevant multiexponential function~\cite{Suter1998}
\begin{equation}
1-\frac{m(t)}{m(\infty)}= \frac{9}{35}e^{\left(-\frac{t}{T_1}\right)} +\frac{4}{15}e^{\left(-\frac{6t}{T_1}\right)} +\frac{10}{21}e^{\left(-\frac{15t}{T_1}\right)}.
\end{equation}
As explained in~\cite{Gavilano1995}, the spin-lattice relaxation is rather complicated at ambient pressure (and also at 0.4 kbar). Here we restrict ourselves to pressures at and above 1.25 kbar, where the ``correlated phase'' is suppressed (see fig.~\ref{SPC_all}). The fraction of powder grains, where the quadrupolar signal is within the irradiation window, is less than 15\%. This leads to a correction of the above recovery function and to the values of $T_1$, in the form of a reduction of maximally 20\%. In figure~\ref{T1T}, the low temperature relaxation rates $T_1^{-1}$ divided by the temperature $T$ are plotted versus $T$ for various pressures. Shown for comparison are the ambient pressure data that were reported in ref.~\cite{Gavilano1995}. The $'\times'$ symbols are relaxation rates from the spectral part between 0.9 and 1.3~kG, reflecting the paramagnetic phase and the $'+'$ symbols represent data points obtained from the broadest part of the ambient pressure spectrum (see figure~\ref{T1T}).
Most obvious is the significant reduction of $(T_1T)^{-1}$ with increasing pressure by almost 2 orders of magnitude in the covered pressure range. This signals a diminishing density of electronic states at the Fermi energy and is in qualitative agreement with the pressure induced reduction of the Knight shift components that results from the analysis of the $^{27}$Al NMR spectra. The temperature variation of $(T_1T)^{-1}(T)$ is fairly small for non zero pressures. Considering the spectra in fig.~\ref{SPC_all} and the $T_1^{-1}$ data in fig.~\ref{T1T} it may be noted that the pressure-induced single phase does not order magnetically above 50 mK. This and the reduction of $D(E_F)$ are both consistent with previously published specific-heat data obtained at high pressures~\cite{Brodale1986}. In table~\ref{ceal3_summarytable} we list approximate values of $(T_1T)^{-1}$, $T_1TK_{\rm iso}^2$ and the parameter $R$ which appears in the well known relation~\cite{Slichter1996}
\begin{equation}
T_1TK^2R=\frac{\hbar}{4\pi k_B}\left(\frac{\gamma_e}{\gamma_N}\right)^2 .
\end{equation}
The parameter $R$ reflects the ratio between the effective magnetic susceptibility $\chi$ and density of electronic states $D(E_F)$. The decrease of $R$ must therefore be traced back to a strong reduction of $\chi(q)$. A previously published neutron scattering study on La-doped CeAl$_3$~\cite{Goremychkin2000} also indicated that pressure diminishes the tendency to a magnetic instability in that material at low temperatures.
We emphasize that our main claims do not depend on the results of the details of the analysis. They are evident from the raw data. Nevertheless, also the quoted pressure-induced variations of the relevant parameters resulting from our quantitative analysis are significant.
\begin{table}
\begin{center}
\begin{tabular}{cccc}
\hline\hline
~$p$~ 	& ~$(T_1T)^{-1}$~ 	& ~$T_1TK_{\rm iso}^2$~ & ~$R$~  \\ 
~kbar~ 	& ~s$^{-1}$K$^{-1}$~ 	& ~$10^{-6}$ s K~ 	& ~1~  \\ 
\hline
0 	& 68			& -		& -		\\ 
2.17 	& $10\pm2$		& $44\pm10$	& $0.088\pm0.021$	 \\ 
5.6 	& $1.2\pm0.24$		& $83\pm20$	& $0.046\pm0.011$		 \\ 
9.15 	& $0.7\pm0.014$		& $115\pm28$	& $0.033\pm0.008$ 	 \\
\hline\hline
\end{tabular}
\end{center}
\caption{Pressure dependence of relaxation-related NMR parameters for the paramagnetic phase at low temperatures.  }
\label{ceal3_summarytable}
\end{table}

We conclude that in CeAl$_3$ the inhomogeneous ground state is quenched by the application of moderate pressures. The so-called correlated phase, reflected in the broad part of the low-pressure Al NMR spectra, vanishes upon increasing external pressure between 0.42 and 1.25~kbar. The remaining phase is modified in the sense that we observe a clear drop in $D(E_F)$ with increasing pressure, reflected in the SLRR and in the Knight shift, which has also been suggested on the basis of macroscopic quantities such as the specific heat or the electrical resistivity~\cite{Brodale1986, Kagayama1994}.
The observations imply that the inhomogeneous ground state at ambient pressure is indeed an intrinsic property of CeAl$_3$ and that the broad line, previously attributed to the correlated phase, is not simply an artefact. A more detailed account of this work will be published elsewhere.
\begin{acknowledgments}
We thank J. Hinderer, J. Kanter, and H. R. Aeschbach for their help in the preparation of the experiments.
This work was, in part, financially supported by the "Schweizerischer Nationalfonds zur F\"orderung der wissenschaftlichen Forschung".
\end{acknowledgments}


\begin{thebibliography}{17}
\expandafter\ifx\csname natexlab\endcsname\relax\def\natexlab#1{#1}\fi
\expandafter\ifx\csname bibnamefont\endcsname\relax
  \def\bibnamefont#1{#1}\fi
\expandafter\ifx\csname bibfnamefont\endcsname\relax
  \def\bibfnamefont#1{#1}\fi
\expandafter\ifx\csname citenamefont\endcsname\relax
  \def\citenamefont#1{#1}\fi
\expandafter\ifx\csname url\endcsname\relax
  \def\url#1{\texttt{#1}}\fi
\expandafter\ifx\csname urlprefix\endcsname\relax\def\urlprefix{URL }\fi
\providecommand{\bibinfo}[2]{#2}
\providecommand{\eprint}[2][]{\url{#2}}

\bibitem[{\citenamefont{Andres et~al.}(1975)\citenamefont{Andres, Graebner, and
  Ott}}]{Andres1975}
\bibinfo{author}{\bibfnamefont{K.}~\bibnamefont{Andres}},
  \bibinfo{author}{\bibfnamefont{J.~E.} \bibnamefont{Graebner}},
  \bibnamefont{and} \bibinfo{author}{\bibfnamefont{H.~R.} \bibnamefont{Ott}},
  \bibinfo{journal}{Phys. Rev. Lett.} \textbf{\bibinfo{volume}{35}},
  \bibinfo{pages}{1779} (\bibinfo{year}{1975}).

\bibitem[{\citenamefont{Barth et~al.}(1987)\citenamefont{Barth, Ott, Gygax,
  Hitti, Lippelt, Schenck, Baines, van~den Brandt, Konter, and
  Mango}}]{Barth1987}
\bibinfo{author}{\bibfnamefont{S.}~\bibnamefont{Barth}},
  \bibinfo{author}{\bibfnamefont{H.~R.} \bibnamefont{Ott}},
  \bibinfo{author}{\bibfnamefont{F.~N.} \bibnamefont{Gygax}},
  \bibinfo{author}{\bibfnamefont{B.}~\bibnamefont{Hitti}},
  \bibinfo{author}{\bibfnamefont{E.}~\bibnamefont{Lippelt}},
  \bibinfo{author}{\bibfnamefont{A.}~\bibnamefont{Schenck}},
  \bibinfo{author}{\bibfnamefont{C.}~\bibnamefont{Baines}},
  \bibinfo{author}{\bibfnamefont{B.}~\bibnamefont{van~den Brandt}},
  \bibinfo{author}{\bibfnamefont{T.}~\bibnamefont{Konter}}, \bibnamefont{and}
  \bibinfo{author}{\bibfnamefont{S.}~\bibnamefont{Mango}},
  \bibinfo{journal}{Phys. Rev. Lett.} \textbf{\bibinfo{volume}{59}},
  \bibinfo{pages}{2991} (\bibinfo{year}{1987}).

\bibitem[{\citenamefont{Barth et~al.}(1989)\citenamefont{Barth, Ott, Gygax,
  Hitti, Lippelt, Schenck, and Baines}}]{Barth1989}
\bibinfo{author}{\bibfnamefont{S.}~\bibnamefont{Barth}},
  \bibinfo{author}{\bibfnamefont{H.~R.} \bibnamefont{Ott}},
  \bibinfo{author}{\bibfnamefont{F.~N.} \bibnamefont{Gygax}},
  \bibinfo{author}{\bibfnamefont{B.}~\bibnamefont{Hitti}},
  \bibinfo{author}{\bibfnamefont{E.}~\bibnamefont{Lippelt}},
  \bibinfo{author}{\bibfnamefont{A.}~\bibnamefont{Schenck}}, \bibnamefont{and}
  \bibinfo{author}{\bibfnamefont{C.}~\bibnamefont{Baines}},
  \bibinfo{journal}{Phys. Rev. B} \textbf{\bibinfo{volume}{39}},
  \bibinfo{pages}{11695} (\bibinfo{year}{1989}).

\bibitem[{\citenamefont{Gavilano et~al.}(1995)\citenamefont{Gavilano,
  Hunziker, and Ott}}]{Gavilano1995}
\bibinfo{author}{\bibfnamefont{J.~L.} \bibnamefont{Gavilano}},
  \bibinfo{author}{\bibfnamefont{J.}~\bibnamefont{Hunziker}}, \bibnamefont{and}
  \bibinfo{author}{\bibfnamefont{H.~R.} \bibnamefont{Ott}},
  \bibinfo{journal}{{Phys. Rev. B}} \textbf{\bibinfo{volume}{52}},
  \bibinfo{pages}{R13106} (\bibinfo{year}{1995}).

\bibitem[{\citenamefont{Nakamura et~al.}(1988)\citenamefont{Nakamura, Kitaoka,
  Asayama, and Flouquet}}]{Nakamura1988}
\bibinfo{author}{\bibfnamefont{H.}~\bibnamefont{Nakamura}},
  \bibinfo{author}{\bibfnamefont{Y.}~\bibnamefont{Kitaoka}},
  \bibinfo{author}{\bibfnamefont{K.}~\bibnamefont{Asayama}}, \bibnamefont{and}
  \bibinfo{author}{\bibfnamefont{J.}~\bibnamefont{Flouquet}},
  \bibinfo{journal}{J. Magn. Magn. Mat.} \textbf{\bibinfo{volume}{76-77}},
  \bibinfo{pages}{465} (\bibinfo{year}{1988}).

\bibitem[{\citenamefont{Wong and Clark}(1992)}]{Wong1992}
\bibinfo{author}{\bibfnamefont{W.~H.} \bibnamefont{Wong}} \bibnamefont{and}
  \bibinfo{author}{\bibfnamefont{W.~G.} \bibnamefont{Clark}},
  \bibinfo{journal}{J. Magn. Magn. Mat.} \textbf{\bibinfo{volume}{108}},
  \bibinfo{pages}{175} (\bibinfo{year}{1992}).

\bibitem[{\citenamefont{Brodale et~al.}(1986)\citenamefont{Brodale, Fisher,
  Phillips, and Flouquet}}]{Brodale1986}
\bibinfo{author}{\bibfnamefont{G.~E.} \bibnamefont{Brodale}},
  \bibinfo{author}{\bibfnamefont{R.~A.} \bibnamefont{Fisher}},
  \bibinfo{author}{\bibfnamefont{N.~E.} \bibnamefont{Phillips}},
  \bibnamefont{and} \bibinfo{author}{\bibfnamefont{J.}~\bibnamefont{Flouquet}},
  \bibinfo{journal}{Phys. Rev. Lett.} \textbf{\bibinfo{volume}{56}},
  \bibinfo{pages}{390} (\bibinfo{year}{1986}).

\bibitem[{\citenamefont{Kagayama et~al.}(1994)\citenamefont{Kagayama, Ishii,
  and Oomi}}]{Kagayama1994}
\bibinfo{author}{\bibfnamefont{T.}~\bibnamefont{Kagayama}},
  \bibinfo{author}{\bibfnamefont{T.}~\bibnamefont{Ishii}}, \bibnamefont{and}
  \bibinfo{author}{\bibfnamefont{G.}~\bibnamefont{Oomi}}, \bibinfo{journal}{J.
  Alloys \& Compounds} \textbf{\bibinfo{volume}{207-208}}, \bibinfo{pages}{263}
  (\bibinfo{year}{1994}).

\bibitem[{\citenamefont{{Reyes} et~al.}(1992)\citenamefont{{Reyes}, {Ahrens},
  {Heffner}, {Hammel}, and {Thompson}}}]{Reyes1992}
\bibinfo{author}{\bibfnamefont{A.~P.} \bibnamefont{{Reyes}}},
  \bibinfo{author}{\bibfnamefont{E.~T.} \bibnamefont{{Ahrens}}},
  \bibinfo{author}{\bibfnamefont{R.~H.} \bibnamefont{{Heffner}}},
  \bibinfo{author}{\bibfnamefont{P.~C.} \bibnamefont{{Hammel}}},
  \bibnamefont{and} \bibinfo{author}{\bibfnamefont{J.~D.}
  \bibnamefont{{Thompson}}}, \bibinfo{journal}{Rev. Sci. Instr.}
  \textbf{\bibinfo{volume}{63}}, \bibinfo{pages}{3120} (\bibinfo{year}{1992}).

\bibitem{footnote1}{As pressure medium, a Balzers silicone oil for vacuum pumps was used.}

\bibitem[{\citenamefont{Carter et~al.}(1977)\citenamefont{Carter, Bennett, and
  Kahan}}]{Carter1977}
\bibinfo{author}{See, e.g., \bibfnamefont{G.~C.} \bibnamefont{Carter}},
  \bibinfo{author}{\bibfnamefont{L.~H.} \bibnamefont{Bennett}},
  \bibnamefont{and} \bibinfo{author}{\bibfnamefont{D.~J.} \bibnamefont{Kahan}},
  \emph{\bibinfo{title}{{Metallic shifts in NMR: a review of the theory and
  comprehensive critical data compilation of metallic materials}}}, Progress in
  Materials Science (\bibinfo{publisher}{Oxford Pergamon Press},
  \bibinfo{address}{New York}, \bibinfo{year}{1977}).

\bibitem{footnote2}{The anisotropy $\epsilon$ of the Knight shift is not restricted to zero. However, the agreement between the simulations and experiment was found to be worse for $\epsilon>0$.}

\bibitem{footnote3}{The pulse sequence used for SLRR measurements consisted of 10 comb pulses of 8~$\mu$s length and 35~$\mu$s delay, followed by the variable delay. The subsequent spin echo consisted of a 4~$\mu$s-pulse, a delay $\tau=50~\mu$s and a second pulse of 8~$\mu$s duration. Avoiding heating effects was a major concern in this experiment. For instance, using other intermetallic compounds in the same pressure cell,  we measure $T_1$ by employing longer rf-combs with much higher rf-pulse amplitudes and frequencies and otherwise the same conditions. We found that heating effects could be observed only below 0.5~K. By reducing the rf-pulse energies and frequencies and decreasing the number of pulses in the rf comb to levels still higher to those used here, we found that the useful temperature range without noticable heating effects could be extended down to 0.25~K.}

\bibitem[{\citenamefont{Hunziker et~al.}(1996)\citenamefont{}}]{Hunziker1996}
\bibinfo{author}{\bibfnamefont{J.~W.}~\bibnamefont{Hunziker}},
  \bibinfo{thesis}{PhD thesis} \textbf{\bibinfo{DissNo}{Diss. ETH No. 11613}},
   (\bibinfo{year}{1996}).

\bibitem[{\citenamefont{Suter et~al.}(1998)\citenamefont{Suter, Mali, Roos, and
  Brinkmann}}]{Suter1998}
\bibinfo{author}{\bibfnamefont{A.}~\bibnamefont{Suter}},
  \bibinfo{author}{\bibfnamefont{M.}~\bibnamefont{Mali}},
  \bibinfo{author}{\bibfnamefont{J.}~\bibnamefont{Roos}}, \bibnamefont{and}
  \bibinfo{author}{\bibfnamefont{D.}~\bibnamefont{Brinkmann}},
  \bibinfo{journal}{J. Phys.: Condensed Matter} \textbf{\bibinfo{volume}{10}},
  \bibinfo{pages}{5977} (\bibinfo{year}{1998}).

\bibitem[{\citenamefont{Joon et~al.}(1989)\citenamefont{Joon, Heinmaa, and
  Skripov}}]{Joon1989}
\bibinfo{author}{\bibfnamefont{E.~R.} \bibnamefont{Joon}},
  \bibinfo{author}{\bibfnamefont{I.~A.} \bibnamefont{Heinmaa}},
  \bibnamefont{and} \bibinfo{author}{\bibfnamefont{A.~V.}
  \bibnamefont{Skripov}}, \bibinfo{journal}{Solid State Comm.}
  \textbf{\bibinfo{volume}{71}}, \bibinfo{pages}{1061} (\bibinfo{year}{1989}).

\bibitem[{\citenamefont{Slichter}(1996)}]{Slichter1996}
\bibinfo{author}{\bibfnamefont{C.~P.} \bibnamefont{Slichter}},
  \emph{\bibinfo{title}{{Principles of Magnetic Resonance}}}, Springer Series
  in Solid-State Sciences (\bibinfo{publisher}{Springer},
  \bibinfo{address}{Berlin}, \bibinfo{year}{1996}).

\bibitem[{\citenamefont{Goremychkin et~al.}(2000)\citenamefont{Goremychkin,
  Osborn, Rainford, and Murani}}]{Goremychkin2000}
\bibinfo{author}{\bibfnamefont{E.~A.} \bibnamefont{Goremychkin}},
  \bibinfo{author}{\bibfnamefont{R.}~\bibnamefont{Osborn}},
  \bibinfo{author}{\bibfnamefont{B.~D.} \bibnamefont{Rainford}},
  \bibnamefont{and} \bibinfo{author}{\bibfnamefont{A.~P.}
  \bibnamefont{Murani}}, \bibinfo{journal}{Phys. Rev. Lett.}
  \textbf{\bibinfo{volume}{84}}, \bibinfo{pages}{2211} (\bibinfo{year}{2000}).

\end{thebibliography}
\end{document}